\documentclass[12pt]{article}

\newcommand {\slsh} [1] {\not{\hbox{\kern-2pt${#1}$}}}

\usepackage[latin1]{inputenc}
\usepackage[T1]{fontenc}
\usepackage[english]{babel}
\usepackage{amsmath}
\usepackage{amsfonts}
\usepackage{latexsym}
\usepackage{indentfirst}
\usepackage{epic}
\usepackage{graphicx}

\newcommand {\beq} {\begin{equation}}
\newcommand {\eeq} {\end{equation}}
  \newcommand {\ber}{\begin{eqnarray*}}
  \newcommand {\eer} {\end{eqnarray*}}
\newcommand {\bea}{\begin{eqnarray}}
  \newcommand {\eea} {\end{eqnarray}}

\newcommand{\None}{${\cal N}=1\ $}

\newcommand{\Dslash}{\,{\raise.15ex\hbox{/}\mkern-12mu D}}
\newcommand{\Tr}{{\rm Tr}\,}
\newcommand{\tr}{{\rm tr}\,}
\newcommand{\gsim}{\lower.7ex\hbox{$
\;\stackrel{\textstyle>}{\sim}\;$}}
\newcommand{\lsim}{\lower.7ex\hbox{$
\;\stackrel{\textstyle<}{\sim}\;$}}

\def\beqn{\begin{eqnarray}}
\def\eeqn{\end{eqnarray}}

\newcommand{\Pexp}{{\rm Pexp}\ }
\newcommand{\lssf}{{}^{\hspace{.25em}f\hspace{-.25em}}}
\newcommand{\lssb}{{}^{\hspace{.2em}b\hspace{-.2em}}}

\begin{document}
\begin{titlepage}

\vskip 0.5cm

\centerline{{\Large \bf Degeneracy Between the Regge Slope of}}

\vskip 0.2cm

\centerline{{\Large \bf Mesons and Baryons from Supersymmetry}}
\vskip 1cm
\centerline{\large Adi Armoni and Agostino Patella}

\vskip 0.3cm
\centerline{\it Department of Physics, Swansea University,}
\centerline{\it Singleton Park, Swansea, SA2 8PP, UK}

\vskip 1cm

\begin{abstract}

We consider the degeneracy between the Regge slope of mesons and baryons in QCD. We argue that within the ``orientifold large-$N$ approximation'' asymptotically massive mesons and baryons become supersymmetric partners and hence degenerate. To this end, we generalize QCD by a $SU(N)$ theory with a quark in the two-index antisymmetric representation. We show that in this framework the meson is represented by an oriented bosonic QCD-string and the baryon is represented by an un-oriented fermionic QCD-string. At large-$N$, due to an equivalence with super Yang-Mills, the tensions of the bosonic and the fermionic strings coincide. Our description of mesons and baryons as oriented and un-oriented bosonic and fermionic QCD-strings is in full agreement with the spectra of open strings in the dual type 0' string theory.
    
\end{abstract}

\end{titlepage}

\section{Introduction}

Despite of the great success of QCD in describing the strong interaction, the low-energy (infra-red) regime, where quarks and gluons are strongly coupled, remains notoriously difficult. Despite a lack of analytical proof of color confinement, we are confident that quarks are confined by a strong force to form mesons and baryons.

One of the interesting aspects of the hadronic spectrum is that both mesons and baryons fall into a linear Regge trajectory of the form
\beq \label{eq:regge}
J= \alpha _0 + \alpha ' M^2 \, ,
\eeq
where $\alpha' \sim 0.89 \, {\rm GeV}^{-2}$ is called the Regge slope. A Regge trajectory for mesons can be easily explained by string theory (see~\cite{PandoZayas:2003yb} for a recent work). A fastly rotating relativistic string admits the relation $M^2 = 2\pi \sigma J$ and hence an intuitive picture of the Regge phenomenon is that the flux tube that connects the quark anti-quark pair behaves like a string with $\sigma = {1\over 2\pi \alpha'}$. This string is often called
 the QCD-string.

It is less obvious why baryons admit a linear Regge trajectory and why the Regge slope of mesons and baryons is the same. A possible explanation (see a review by Wilczek~\cite{Wilczek:2004im}) is that the baryon structure is similar to the meson structure: the baryon is composed of a quark and a diquark connected by a single long string. It is not clear whether a generalization of this picture applies to $SU(N)$, or whether it holds for $SU(3)$ only.

In this paper we propose a new explanation for the coincidence of the Regge slopes of baryons and mesons. Our derivation is based on ``orientifold planar equivalence''.

``Orientifold planar equivalence'' is the statement that an $SU(N)$ gauge theory with a massless Dirac fermion in the antisymmetric representation (commonly called ``orientifold field theory'') becomes equivalent in a certain well-defined sector to $SU(N)$ \None Super Yang-Mills~\cite{Armoni:2003gp,Armoni:2004ub}. A necessary and sufficient condition for the equivalence is an unbroken charge conjugation symmetry~\cite{Unsal:2006pj}. Provided that the equivalence holds it enables us to copy results from the supersymmetric theory to the non-supersymmetric theory~\cite{Armoni:2003fb}. Moreover, since for $SU(3)$ the fundamental and the anti-symmetric representation are equivalent to each other, planar equivalence provides a way to estimate non-perturbative quantities in QCD by copying them from \None super Yang-Mills. It led to various analytical non-perturbative results in QCD, including a calculation of the quark condensate~\cite{Armoni:2003yv}, see review~\cite{Armoni:2004uu}
 .

The idea of the present paper is to describe the meson and the baryon in the ``orientifold theory'' by operators whose mass spectra become at large-$N$ degenerate via the relation with super Yang-Mills. Briefly, the meson is a quark anti-quark pair connected by a QCD-string and the baryon is a quark quark pair connected by a string which contains an anti-symmetric fermion smeared on it.\footnote{There are other generalizations of baryons in ``orientifold theories'', see~\cite{Bolognesi:2009vm} and references therein.} We show that the two QCD-strings become superpartners at infinite $N$. Therefore, in our picture, the meson and the baryon Regge slopes are degenerate due to supersymmetric relics in QCD~\cite{Armoni:2003fb}. A related, however different, "effective supersymmetry" was assumed in \cite{Catto:1984wi,Lichtenberg:1999sc,Karliner:2006fr} in order to explain spectra degeneracies between mesons and baryons.

Our picture of the baryon is similar to the quark-diquark picture in the sense that there is only one long QCD-string in the baryon. However, in our picture there is a smeared quark along the string, whereas in the quark-diquark picture there is one quark at one end of the string and two quarks at the other end. 

\section{Meson and Baryons in ``Orientifold Field Theories''} 

Consider an $SU(N)$ gauge theory with one massless Dirac fermion $\Psi$ in the two-index antisymmetric representation and one or more Dirac fermions $Q$ in the fundamental representation. Let us consider the following operators (in the Hamiltonian formalism, therefore $A_0=0$; see appendix.~\ref{app:hamiltonian} for a review of the main concepts)
\bea
& & M = \bar Q^{j} (x_0) \left (\Pexp i \int_{x_0}^{x_1} \vec{A}(y) d\vec{y}  \right ) ^i _j Q_{i}(x_1) \label{operators} \\
& & B_\alpha  = \int d\vec{z} \, Q_{j}^C (x_0) \left (\Pexp i \int_{x_0}^z \vec{A}(y) d\vec{y}  \right ) ^j _k \vec{\gamma}_{\alpha\beta} \bar\Psi ^{[kl]}_\beta (z) \left (\Pexp i \int_z^{x_1} \vec{A}(y) d\vec{y} \right ) ^i _l Q_{i} (x_1) \,  . \nonumber
\eea
We assume the same path connecting $x$ and $y$ for both $M$ and $B$. $z$ is an intermediate point along this path. At any $N$, $M$ is a bosonic operator and $B$ is a fermionic operator ($\alpha$ is its spin index). We define $Q^C = Q^T C$ where $C$ is the charge-conjugation matrix acting on the spin index. Assigning to $Q$ and $\Psi$ respectively baryonic numbers $1/3$ and $-1/3$, the operator $M$ create a mesonic state (though it is an eigenvalue neither of the Hamiltonian nor of the angular momentum), while the operator $B$ creates a baryonic state, built with two quarks $Q$ and a fermion $\bar{\Psi}$. 
We think about $M$ and $B$ as a generalization of ordinary mesons and baryons in QCD. Indeed since for $SU(3)$ $q_i = {1\over 2} \epsilon_{ijk} \bar \Psi^{[jk]}$, the fermion $\bar{\Psi}$ can be equivalently replaced by one more quark in the fundamental representation, and the baryonic number defined above is nothing but the usual baryonic number of QCD. This generalization of baryons was firstly proposed by Corrigan and Ramond in ref.~\cite{Corrigan:1979xf}. 

It is well known that at high angular momentum (but below the string-breaking energy), the mesons are effectively described by a relativistic semiclassical quark-antiquark pair, connected by a bosonic string. The Regge slope of mesons made of light quarks is related to the tension $\sigma_b$ of the bosonic string (see for instance~\cite{Baker:2002km}):
\beq
\alpha'_M = \frac{1}{2\pi \sigma_b} \, .
\eeq
In a theory with $\Psi$ fermions, a fermionic string is present as well. Microscopically it can be thought as a chromoelectric flux with a $\Psi$ fermion smeared in it. In analogy with mesons, baryons of the kind generated by the operator $B$ are effectively described by a relativistic semiclassical quark-quark pair, connected by a fermionic string. This description implies a Regge trajectory for baryons made of light quarks, and the slope is related to the tension $\sigma_f$ of the fermionic string:
\beq
\alpha'_B = \frac{1}{2\pi \sigma_f} \, .
\eeq
We want to show that in the large-$N$ limit the two string tensions are equal, proving this way the equality of the Regge slopes.

The strategy is the following. We consider infinitely heavy quarks $Q$ and we use them to probe the string states. Even if mesons and baryons made of heavy quarks do not follow the Regge trajectory in the form~\eqref{eq:regge} (but the mass gets a shift~\cite{Baker:2002km}), we will be able in this sector to get the degeneracy of the bosonic and fermionic string tensions. Light quarks $Q$ do not spoil the degeneracy at large $N$ of the string tensions. In fact charges in the fundamental representation are quenched in the large-$N$ limit, therefore the string tensions do not depend on the mass of the quarks $Q$.

\vskip 1cm

Assuming then heavy quarks $Q$, we decompose the states obtained with the operators $M$ and $B$ acting on the vacuum in eigenstates of the Hamiltonian (see appendix~\ref{app:static} for the definition of these states). At large separation $R$ of the heavy quarks, the energy states are proportional to $R$:
\bea
&& M \left| 0 \right> = \sum_{n r} M_{n r} \left| n, r \right>^b \\
&& H \left| n, r \right>^b = \sigma^b_n R \left| n, r \right>^b + O(R^0) \\
&& B_\alpha \left| 0 \right> = \sum_{n r} B_{\alpha n r} \left| n,r \right>^f \\
&& H \left| n, r \right>^f = \sigma^f_n R \left| n, r \right>^f + O(R^0)
\eea
where the superscripts $b/f$ stand for bosonic or fermionic and $r$ is an index running on all the degenerate states (if any) in the same energy level $n$. The set of states $\left| n,r \right>^{b/f}$ are supposed to be an orthonormal basis of the Hilbert subspace describing the system in presence of two static charges in $x_0$ and $x_1$. Since for $SU(3)$ this theory is ordinary QCD, we will refer to the sets $\{ \sigma^b_n \}$ and $\{ \sigma^f_n \}$ respectively as the spectra of the bosonic and fermionic QCD-strings.

Let us discuss the relation with ${\cal N}=1$ super Yang-Mills. We replace in \eqref{operators} the antisymmetric fermion by an adjoint fermion, and some heavy quarks by heavy anti-quarks where necessary 
\bea
& & \tilde M = \bar Q^j (x_0) \left (\Pexp i \int_{x_0}^{x_1} \vec{A}(y) d\vec{y}  \right ) ^i _j Q_i (x_1) \, \label{operators2}  \\ 
& & \tilde B_\alpha = \int d\vec{z} \, \bar Q^j (x_0) \left (\Pexp i \int_{x_0}^z \vec{A}(y) d\vec{y}  \right )_j^k \vec{\gamma}_{\alpha\beta} \lambda_{k\beta}^l(z) \left (\Pexp i \int_z^{x_1} \vec{A}(y) d\vec{y} \right ) ^i _l Q_i  (x_1) \, . \nonumber 
\eea
As we will see later on, planar equivalence implies that at large-$N$ the spectrum of the QCD-strings in the orientifold theory coincides with the spectrum of the effective strings in super Yang-Mills, which we show to be boson/fermion degenerate.

The result is not obvious because the presence of the static charges explicitly breaks SUSY, since we do not introduce heavy squarks as well. We will argue, however, that the spin of the charges at the end-points of the QCD-string should not affect its tension. In particular, having only quarks (and not squarks) at the end-points will not spoil the boson/fermion degeneracy. 

If we consider SUSY transformations in the de Wit-Freedman form~\cite{de Wit:1975nq}, in the Hamiltonian gauge and without squarks ($S_\alpha$ is the supercharge in the 4-component spinor notation)
\bea
&& [S, A^a_k] = \gamma_k \lambda^a \\
&& [S_\alpha, \lambda^a_\beta] = -\frac{1}{4} F^a_{\mu \nu} [\gamma^\mu, \gamma^\nu]_{\alpha\beta} \\
&& [S_\alpha, Q_{i\alpha}] = 0 \\
&& \sum_\alpha (S^\dagger)^\alpha S_\alpha = 4 H_{SYM} \label{eq:charge2}
\eea
then the variation of the Hamiltonian is in general not zero, but is proportional to the chromoelectric charge density of the heavy quarks (see appendix~\ref{app:SYM} for more details):
\beq
[S, H_{SYM}] = \int d^3x \ \gamma_0 \lambda^a Q^\dagger t^a Q \, . \label{eq:basic_commutator}
\eeq

Chosen a bosonic state $\left| n, r \right>^b$, from equation~\eqref{eq:charge2} we get:
\beq
4 \sigma^b_n R = \sum_\alpha \lssb\left<n,r \right| (S^\dagger)^\alpha S_\alpha \left| n, r \right>^b = \sum_{\alpha,n',r'} \left| \lssf\left<n',r' \right| S_\alpha \left| n, r \right>^b \right|^2
\eeq
which means that at least one fermionic state $\left| n', r' \right>^f $ and one spin component $\alpha$ exist with the following property:
\beq
\lssf\left< n',r' \right| S_\alpha \left| n, r \right>^b = O\left(\sqrt{R}\right) \, .
\eeq
Now we can compute the element matrix of the commutator $[S_\alpha,H]$ between this pair of states, using equation~\eqref{eq:basic_commutator}
\beq
( \sigma^b_n - \sigma^f_{n'} ) \lssf\left< n',r' \right| S_\alpha \left| n, r \right>^b =
{1\over R} \lssf\left< n',r' \right| \int d^3x \ (\gamma_0 \lambda^a)_\alpha Q^\dagger t^a Q \left| n, r \right>^b \, . \label{delta}
\eeq
Being the chromoelectric charge density localised only in the points $x_0$ and $x_1$, we expect that at large charge separation, $R\rightarrow \infty $, the r.h.s. of equation \eqref{delta} will vanish. The reason is that the charge density is not expected to give rise to a linear dependence on $R$.
As a result,
\beq
\sigma^b_n = \sigma^f_{n'} \, .
\eeq
At large $R$ the energy levels are Fermi/Bose degenerate and we can reorganise the numeration of the fermionic levels requiring that $n'=n$. The formalism above translates in mathematical terms the idea that SUSY is broken only by boundary terms and therefore is recovered in the limit of large charge separation $R$.

The operators $\tilde M$ and $\tilde B$ are connected by a SUSY transformation
\beq
[S_\alpha, \tilde M] =  \tilde B_\alpha \, .
\eeq
This implies that the operator $\tilde B_\alpha$ creates a baryon in an energy level $n$ with non-vanishing amplitude if and only if the operator $\tilde M$ creates a meson in the same energy level (at the leading order in $R$) with non-vanishing amplitude.

By copying the above result to the ``orientifold theory'' we conclude that the QCD-string tension of the meson is identical to the QCD-string tension of the baryon. In other words the Regge slope of mesons and baryons is identical at large-$N$. Another observation is that for $SU(2)$ the antisymmetric fermion becomes a singlet and therefore in this limit the mass spectra of $M$ and $B$ become degenerate. For this reason we expect that the $1/N$ corrections will not be large and therefore the large-$N$ result should be a good approximation for QCD.

It is possible to re-formulate the above discussion in terms of Polyakov loops. Despite of the equivalence of the two formulations, the derivation via Polyakov loops will enable an easier connection with type 0' string theory. We will also able to see how orientifold planar equivalence can be extended in the string sector.

We consider the orientifold theory on the Euclidean space $R^3 \times S^1$. We assume that the circle is large and that the theory is in the confining phase, namely that the expectation value of a Polyakov loop that wraps the circle is zero $\langle P(x) \rangle =0$.

Consider the two-point function 
\beq 
\langle {\rm Re} \, P(x) \, {\rm Re} \, P(y) \rangle \, , \label{correlator}
\eeq
where $x,y$ are coordinates in $R^3$.

Being ${\rm Re} \, P$ a bosonic operator invariant under charge conjugation (i.e. it belongs to the neutral sector), by orientifold planar equivalence the above correlator \eqref{correlator} is equal to the same correlator in pure \None Super Yang-Mills~\cite{Unsal:2006pj}, and expanding the real part we get
\beq
 \langle  P(x) \, P(y) \rangle _{\rm Orienti} +
\langle  P(x) \, P^\dagger (y) \rangle _{\rm Orienti}
= 
\langle  P(x) \, P^\dagger (y) \rangle _{\rm SUSY} \label{equivalence} \, .
\eeq
We used the charge conjugation invariance of both the theories, and the invariance of the \None SYM under the global $Z_N$ center symmetry, which implies that the correlator $\langle  P(x)  \, P(y) \rangle _{\rm SUSY}$ identically vanishes. In the orientifold theory the fermions respect only a $Z_2$ global symmetry~\cite{DelDebbio:2008ur} hence $\langle  P(x) \, P(y) \rangle _{\rm Orienti}$ does not need to vanish.

The correlator $\langle  P(x) \, P^\dagger (y) \rangle _{\rm Orienti}$  is saturated by open QCD-strings with a heavy quark and a heavy anti-quark at their ends, namely by {\em oriented} open strings
\beq
  \langle  P(x) \, P^\dagger (y) \rangle _{\rm Orienti}=\sum_n A^b_n \exp - \beta \sigma^b_n |x-y| \, .
\eeq
The correlator $\langle  P(x) \, P(y) \rangle _{\rm Orienti}$ is saturated by open strings which contain heavy quarks at both ends. Thus these string are {\em unoriented}
\beq
  \langle  P(x) \, P(y) \rangle _{\rm Orienti}=\sum _n A^f_n \exp - \beta \sigma^f_n |x-y| \, .
\eeq
The string spectra of the orientifold theory is depicted in figure \eqref{strings} below. We identify the oriented string (which contains a quark and an anti-quark at its ends) with the meson string and the unoriented string (which contain quarks at both ends) with the baryon string of \eqref{operators}.

\begin{figure}[h]
 \centerline{\includegraphics[width=3in]{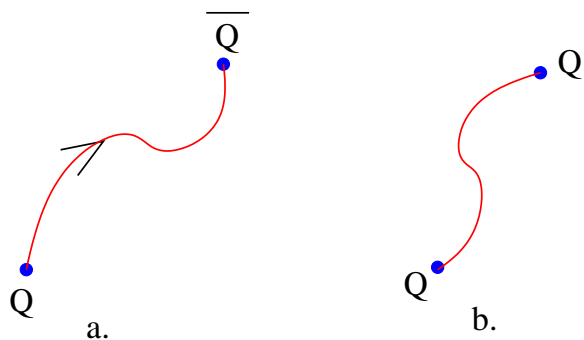}}
 \caption{The open QCD-string spectra of the orientifold theory. a. Oriented (bosonic) strings. b. Un-oriented (fermionic) strings. }
 \label{strings}
 \end{figure}

In \None SYM the correlator $\langle  P(x) \, P^\dagger (y) \rangle _{\rm SUSY}$ is saturated by oriented strings. These strings are both bosonic and fermionic. In the large N limit the orientifold planar equivalence~\eqref{equivalence} gives
\beq
\langle  P(x) \, P^\dagger (y) \rangle _{\rm SUSY}
=\sum _n A^b_n \exp - \beta \sigma^b_n |x-y| 
+\sum _n A^f_n \exp - \beta \sigma^f_n |x-y|
\, .
\eeq
and finally supersymmetry implies that
\beq
\sigma^b_n  = \sigma^f_n \, .
\eeq
which means that the large-$N$ orientifold theory exhibits a degeneracy between the bosonic and the fermionic strings.

The above analysis is in perfect agreement with Sagnotti's model (type 0' string theory)~\cite{Sagnotti:1995ga,Sagnotti:1996qj}. The $SU(N)$ gauge theory with an antisymmetric (or symmetric) fermion lives on a stack of
D-branes in type 0' string theory. The open string spectrum contains both bosons and fermions. The bosonic open strings are oriented and the fermionic open strings are unoriented. Although this string theory is non-supersymmetric, the spectra of open strings exhibits a Bose-Fermi degeneracy.

\section{Conclusions}

In conclusion, we showed that we can generalize QCD in such a way that both the mesons and the baryons are represented by an open string. The meson is a bosonic oriented string and the baryon is a fermionic un-oriented string. The two kinds of open strings admit the same Regge slope. Our assumptions were mild: we assumed that planar equivalence holds and that the color-singlet spectrum of \None super Yang-Mills contains stringy like objects. It will be interesting to perform a lattice simulation which will enable to confirm our result and to estimate the $1/N$ corrections. Finally, it will be interesting to explore the relation between our approach and ``effective supersymmetry'' \cite{Catto:1984wi,Lichtenberg:1999sc,Karliner:2006fr}.

\section*{Acknowledgments}

 A.A.~is supported by the PPARC advanced fellowship award. We thank O. Aharony, O. Bergman, B. Lucini and especially to M. Shifman for discussions.

\appendix

\section{Review of gauge theories in the Hamiltonian formalism}
\label{app:hamiltonian}

In this Appendix we want to review the main concepts, which are useful in this work, about $SU(N)$ gauge theories in the Hamiltonian formalism. The main ingredients are:
\begin{itemize}
\item the Hilbert space defined as the space of the wave functions over the classical field configurations;
\item the Gauss constrain which restricts the physical subspace of the Hilbert space, by requiring invariance under local gauge transformations;
\item the Hamiltonian which defines the dynamics of the system.
\end{itemize}
All the formulae which follow make sense only once a regularization and renormalization scheme is fixed. However for sake of simplicity, the regularization will be kept implicit.

The canonical degrees of freedom are the spatial gauge field $A^A_i$ (the Hamiltonian gauge $A_0=0$ is assumed), the chromoelectric field $E^A_i$, and the matter field $\psi$ (which we will assume in the generic representation $\mathcal{R}$ of the gauge group). The canonical variables satisfy the following (anti)commutation relationships:
\begin{gather}
[E^A_i(x),A^B_j(y)] = - i g^2 \delta^{AB} \delta_{ij} \delta^3(x-y) \, , \\
\{ \psi^a(x), \bar{\psi}^b(y) \} = \gamma^0 \delta^{ab} \delta^3(x-y) \, ,
\end{gather}
where $a$, $b$ are the colour indexed in the representation $\mathcal{R}$.

The Hamiltonian is given by:
\begin{equation} \label{eq:app:hamiltonian}
H = \int \left[ \frac{1}{2g^2} E^A_i E^A_i + \frac{1}{2g^2} B^A_i B^A_i + i \bar{\psi} \gamma_i D_i^{\mathcal{R}} \psi \right] d^3x \, ,
\end{equation}
where $B^A_i = \frac{1}{2} \epsilon_{ijk}F^A_{jk}$ is the chromomagnetic field, and $D_i^{\mathcal{R}}=\partial_i + iT^A_{\mathcal{R}} A^A_i$ is the covariant derivative.

The Hamiltonian is invariant under local gauge transformations, whose generators are:
\begin{equation}
G^A(x) = \frac{1}{g^2} D_i E^a_i(x) + \psi^\dagger T^A_{\mathcal{R}} \psi(x) \, .
\end{equation}
The physical states are defined to be invariant under local gauge transformations, and this condition is the Gauss constrain without any external source:
\begin{equation}
G^A \left| \psi_{phys} \right> = 0 \, .
\label{eq:gauss_no_source}
\end{equation}

The partition function is given by the trace of $e^{-\beta H}$, discarding the non-physical states. This can be achieved by introducing the projector onto the physical states; it is defined as:
\begin{equation}
\mathbb{P}_{0} = \int \exp \left\{ -i \int G^A(x) \omega^A(x) d^3x \right\} \mathcal{D}\Omega[\omega] \, ,
\end{equation}
where $ \exp \{ -i \int G^A(x) \omega^A(x) d^3x \} $ is the operator representing the gauge transformation $\Omega[\omega] = \exp \{ -i T^A \omega^A \}$ on the Hilbert space, and the Haar measure projects on the singlet states. In the path integral formalism, $\omega^A(x)$ which plays the role of a Lagrangian multiplier in the Hamiltonian formalism becomes the $A_O(x)$ component of the gauge field, while $\Omega(x)$ is the Polyakov loop. Finally the partition function is:
\begin{equation}
Z = \Tr \left[ \mathbb{P}_{0} e^{- \beta H} \right] \, .
\end{equation}

\section{Static heavy fundamental charges}
\label{app:static}

In this Appendix we consider a gauge theory with dynamical matter, coupled with two static heavy charges in the fundamental or antifundamental representations, which we will call respectively quark and antiquark. We start with a static heavy quark in the spatial point $x_1$, and a static heavy antiquark in $x_2$. The interaction Hamiltonian is given by:
\begin{equation}
H_{int} = - \int A^A_i \bar{Q} \gamma_i T^A Q d^3x \, ,
\end{equation}
but, since the heavy quarks are static, the current $\bar{Q} \gamma_i T^A Q$ is zero, and the dynamical sector evolves with the same Hamiltonian~\eqref{eq:app:hamiltonian}, as it were not coupled to any static charges. However the quarks modify the Gauss law. The physical states of the whole system (heavy charges plus dynamical degrees of freedom) must be invariant under local gauge transformation. This means that the states of the dynamical sector must transform as the antifundamental representation under gauge transformations in $x_1$, and as the fundamental representation under gauge transformations in $x_2$. We will refer to these, as the states of the effective oriented string connecting the two heavy charges. The projector on the space of the oriented string states is:
\begin{equation}
\mathbb{P}_{Q\bar{Q}}(x_1,x_2) = \int \exp \left\{ i \int G^A(x) \omega^A(x) d^3 x \right\} \tr \Omega(x_1) \tr \Omega(x_2)^\dagger \mathcal{D}\Omega[\omega] \, .
\end{equation}
The last equation essentially comes from the orthogonality of the characters with respect to the integration with the Haar measure. The partition function of the gauge theory in presence of the two heavy charges is therefore given by:
\begin{equation}
Z_{Q\bar{Q}}(\beta,|x_1-x_2|) = \Tr \left[ \mathbb{P}_{Q\bar{Q}}(x_1,x_2) e^{- \beta H} \right] \, .
\end{equation}

In the path integral formalism, the partition function $Z_{Q\bar{Q}}(\beta,|x_1-x_2|)$ is proportional to the expectation value of two Polyakov loops $\Omega(x)$ in the appropriate representations:
\begin{equation} \label{eq:z_pol}
Z_{Q\bar{Q}}(\beta,|x_1-x_2|) = Z(\beta) \langle \tr \Omega(x_1) \tr \Omega(x_2)^\dagger \rangle \, .
\end{equation}

Since the Hamiltonian commutes with the gauge transformations, it can be diagonalized on the space of the string states. In case of dynamical matter in the adjoint representation, there are both bosonic and fermionic states. This can be easily understood considering that an arbitrary numbers of gluinos can be inserted in the flux tube; an even number of gluinos gives rise to bosonic states, while an odd number of gluinos gives rise to fermionic states. The eigenvalues are functions of the distance $R=|x_1-x_2|$. The bosonic eigenstates $\left| n, r \right>^b$ have eigenvalue $V^b_{n,r}(R)$; and the fermionic eigenstates $\left| n, r \right>^f$ have eigenvalue $V^f_{n,r}(R)$:
\begin{gather}
H \left| n, r \right>^b = V^b_{n,r}(R) \left| n, r \right>^b \, , \\
H \left| n, r \right>^f = V^f_{n,r}(R) \left| n, r \right>^f \, .
\end{gather}
The eigenvalues are organized with respect to their large distance behaviour:
\begin{gather}
V^b_{n,r}(R) = \sigma^b_n R + \mathcal{O}(R^0) \, , \\
V^f_{n,r}(R) = \sigma^f_n R + \mathcal{O}(R^0) \, .
\end{gather}
By expanding the trace of the partition function in Eq.~\eqref{eq:z_pol} with respect this basis of states we get for the correlators of Polyakov loops:
\begin{equation} 
\langle \tr \Omega(x_1) \tr \Omega(x_2)^\dagger \rangle_{SYM} = 
\frac{1}{Z_{SYM}(\beta)} \left\{ \sum_{n,r} e^{- \beta V^b_{n,r}(R)} + \sum_{n,r} e^{- \beta V^f_{n,r}(R)} \right\} \, .
\end{equation}

In the case of dynamical matter in the antisymmetric representation, for gauge invariance only an even number of dynamical fermions can be inserted in the flux tube connecting a $Q\bar{Q}$ pair. This means that the oriented string states in the orientifold theory are purely bosonic.
\begin{equation} 
\langle \tr \Omega(x_1) \tr \Omega(x_2)^\dagger \rangle_{Orienti} = 
\frac{1}{Z_{Orienti}(\beta)} \sum_{n,r} e^{- \beta V^b_{n,r}(R)} \, .
\end{equation}
However in the orientifold theory, string states exist connecting a static heavy quark in $x_1$, and another static heavy quark in $x_2$. We will refer to these as the states of the unoriented string. The projector onto the unoriented string states is:
\begin{equation}
\mathbb{P}_{QQ}(x_1,x_2) = \int \exp \left\{ i \int G^A(x) \omega^A(x) d^3 x \right\} \tr \Omega(x_1) \tr \Omega(x_2) \mathcal{D}\Omega[\omega] \, .
\end{equation}
For gauge invariance an odd number of dynamical fermions must be inserted in the flux tube to connect a $QQ$ pair. This means that the unoriented string states in the orientifold theory are purely fermionic. Following the same construction as above, we get:
\begin{equation} 
\langle \tr \Omega(x_1) \tr \Omega(x_2) \rangle_{Orienti} = 
\frac{1}{Z_{Orienti}(\beta)} \sum_{n,r} \sum_{n,r} e^{- \beta V^f_{n,r}(R)} \, .
\end{equation}

\section{SYM in the Hamiltonian formalism}
\label{app:SYM}

In the case of SYM, the formulae of Appendix A must be generalized to the case of a Majorana fermion $\lambda^A$ in the adjoint representation of the gauge group. Using four-component spinors, the Majorana fermion obeys the constrain $\lambda^A_\alpha=C_{\alpha \beta} \bar{\lambda}^A_\beta$ where $C$ is the charge conjugation matrix. With the normalization given by:
\begin{equation}
\{ \lambda^A(x), \bar{\lambda}^B(y) \} = g^2 \gamma^0 \delta^{AB} \delta^3(x-y) \, ,
\end{equation}
the SYM Hamiltonian is:
\begin{equation}
H = \int \left[ \frac{1}{2g^2} E^A_i E^A_i + \frac{1}{2g^2} B^A_i B^A_i + \frac{i}{2g^2} \bar{\lambda}^A \gamma_i D_i \lambda^A \right] d^3x \, ,
\end{equation}
and the generator of the local gauge transformation is:
\begin{equation}
G^A(x) = \frac{1}{g^2} D_i E^a_i(x) + \frac{1}{2g^2} \lambda^\dagger T^A \lambda(x) \, .
\end{equation}

The supersymmetric charge can be introduced in the (on-shell) de Wit-Freedman form:
\begin{gather}
[S, A^A_i(x)] = \gamma_i \lambda^A(x) \, , \\
\{ S_\alpha, \lambda^A_\beta(x) \} = -\frac{1}{4} F_{\mu \nu}^A(x) [\gamma^\mu,\gamma^\nu]_{\alpha \beta} \, .
\end{gather}
In terms of the elementary fields the supercharge is:
\begin{equation}
S = \int \left\{
i E_k^A \gamma_k + B_k^A \gamma_5 \gamma_k
\right\} \lambda^A \ d^3x \, .
\end{equation}
It obeys the Majorana condition $S_\alpha = C_{\alpha \beta} \bar{S}_\beta$, is invariant under spatial translations and local gauge transformations. It does not commute with the Hamiltonian, but:
\begin{equation}
[S,H] = \int \gamma^0 \lambda^A G^A \ d^3x \, .
\end{equation}
It is worth to remind at this point that we are assuming that the theory is properly regularized and renormalized. The equation above is valid only if a regularization scheme which preserves SUSY is chosen, otherwise the commutation relationship gets an extra SUSY violating term, which must vanish as the cutoff is removed.

On the physical states, the supercharge commutes with the Hamiltonian as expected thanks to the Gauss constrain:
\begin{equation}
[S,H] \left| \psi_{phys} \right> = 0 \, ,
\end{equation}
while this is not true for instance on the string states.

Moreover the anticommutator of two supercharges is given by:
\begin{equation}
\{ S, \bar{S} \} = 2 ( \gamma^0 H - \gamma_k \Pi_k ) \, ,
\end{equation}
where $\Pi_k$ are the gauge invariant generalization of the spatial momenta $P_i$:
\begin{equation}
\Pi_k = \int \{ \epsilon_{kij} E_i^A B_j^A + \frac{i}{2} \bar{\lambda}^A \gamma^0 D_k\lambda^A \} \ d^3x = P_k + \int A_k^A G^A \ d^3x \, .
\end{equation}
Again, on the physical states the operators $\Pi_i$ and $P_i$ are identical thanks to the Gauss law.

\end{document}